\documentclass[prb,aps,groupedaddress,twocolumn,floatfix,showpacs]{revtex4}

\usepackage[dvips]{graphicx}
\usepackage{dcolumn}
\usepackage{bm}
\usepackage{times}
\usepackage{amsmath}
\usepackage{epsfig}
\usepackage{dcolumn}
\usepackage{amssymb}
\usepackage{color}

\newcommand{\vc}[1]{{\bf #1}}

\begin{document}

%%%%%%%%%%%%%%%%%%

\title{ Improper s-wave symmetry for the electronic pairing in iron-based
    superconductors by first-principles calculations } 

\author{M. Casula}
\email[]{michele.casula@impmc.upmc.fr}
\affiliation{CNRS and Institut de Min\'eralogie et 
de Physique des Milieux condens\'es,
Universit\'e Pierre et Marie Curie,
case 115, 4 place Jussieu, 75252, Paris cedex 05, France}
\author{S. Sorella}
\email[]{sorella@sissa.it}
\affiliation{International School for Advanced Studies (SISSA) Via Beirut 2,4
  34014 Trieste, Italy and INFM Democritos National Simulation Center,
  Trieste, Italy} 
\date{\today}

\begin{abstract}
By means of space-group symmetry arguments,
we argue that the electronic pairing in iron-based high temperature superconductors 
shows a structure which is a linear combination of \emph{planar}
s-wave and d-wave symmetry channels, both preserving 
the 3-dimensional $A_{1g}$ irreducible representation of the
corresponding crystal point-group. We demonstrate that the s- and
d-wave channels are determined by the parity under reflection of the electronic
orbitals through the iron planes, and by 
improper rotations around the iron sites.
We provide evidence of these general properties by performing accurate quantum Monte Carlo
ab-initio calculations of the pairing function,
for a FeSe lattice with
tetragonal experimental geometry at ambient pressure. 
We find that this picture
survives even in the FeSe under pressure and at low temperatures, when the tetragonal point-group
symmetry is slightly broken.
In order to achieve a higher resolution in momentum space we introduce a BCS model 
that faithfully describes our QMC variational pairing function on the
simulated 4x4 FeSe unit cell.
This allows us to provide a k-resolved image of the pairing function,
and show that non-isotropic contributions in the BCS gap function are
related to the improper s-wave symmetry.
Our theory can rationalize and
explain a series of contradictory experimental findings,
such as the observation of twofold symmetry in the FeSe
superconducting phase,
the anomalous drop of $T_c$ with Co-impurity in 
LaFeAsO$_{(1-x)}$F$_x$, the $s$-to-$d$-wave gap 
transition in BaFe$_2$As$_2$ under K doping, and the nodes appearing in 
the LiFeAs superconducting gap upon P isovalent substitution.
\end{abstract}


\maketitle

\section{Introduction}
\label{intro}

The pairing symmetry of the superconducting state
in iron-based superconductors (IBS) has been one of the most debated subjects since
their first discovery, in both theory and experiments.\cite{paglione,stewart,hirschfeld}
Its determination is particularly challenging in the IBS for their
complex electronic structure,
with a strong multiband character, and a Fermi surface
constituted by many sheets, which can vary with doping and chemical
composition. Moreover, the IBS families are usually compensated
metals, and thus both electrons and holes are involved in the
paired state. 
Weak-coupling RPA approaches, coupled to 
multiband BCS theory, 
yield a pairing function
with global s-wave ($A_{1g}$) symmetry,
but with electron and holes pockets having opposite sign. This scenario, dubbed
$s^\pm$, was first proposed by Mazin 
\emph{et al.}\cite{mazin} complemented later with 
its variants, called ``extended $s^\pm$'',\cite{kuroki2} in the case that accidental
nodes appear on a single Fermi sheet without 
breaking the full rotational symmetry. The latest generalizations
include also ``weak'' nodal lines which develop as closed loops on the
3-dimensional (3D) Fermi surface.\cite{graser2,suzuki,mazin2}

A variety of experiments has been performed to probe the pairing
symmetry of IBS, ranging from thermal conductivity and
specific heat measurements to angle-resolved photoemission
spectroscopy. While there is no doubt on the 
spin singlet nature of the pairing state as revealed by the Knight shift,\cite{ning} 
the presence of nodes and the total symmetry of the spatial part of the
pairing function are controversial. In fact, the experimental outcome
seems to lack universality, as it depends crucially on the ``family'' of
tested compound, its doping, its isovalent substitutions, and its
level of disorder. For instance, the situation is paradoxical for
the ``111'' family, where the LiFeAs material shows a fully gapped
superconducting order parameter, while the isovalent substitution of arsenic by
phosphorus leads to thermodynamic properties compatible with a nodal
pairing function.\cite{hashimoto} 
For the ``122'' family, there
is a recent claim, supported by independent experimental probes,
that the pairing in the BaFe$_2$As$_2$ undergoes an $s$-to-$d$ symmetry
change by doping with potassium,\cite{s2d} the fully substituted
KFe$_2$As$_2$ being identified as a $d$-wave superconductor.\cite{dong} In the
``1111'' family, the dependence of the critical
temperature $T_c$ upon disorder has been studied in the
LaFeAsO$_{(1-x)}$F$_x$  with $x=0.11$. It has been found that Co-doping 
induced disorder makes $T_c$ to fall much more slowly than
Mn-doping.\cite{impjap} 
This 
is certainly not  compatible with a doping independent pairing function with sign changes 
(either $s^\pm$ or $d$-wave).  

 Finally,
specific heat measurements performed on the FeSe 
revealed a highly anisotropic order parameter
in the ``11'' family,\cite{zeng} with a twofold symmetry
directly observed by scanning tunneling microscopy (STM).\cite{twofold}

In this paper, we 
study the IBS gap structure
from a different
perspective, namely 
by looking for 
its
\emph{universal} features based on symmetry constraints induced by
the 3D \emph{space-group} transformations of the crystals. 
We prove that the IBS pairing function is a linear combination
of terms having planar s- and d-wave symmetries, both fulfilling the
full 3D $A_{1g}$ representation.
These properties are
then verified in the FeSe by performing state-of-the-art quantum Monte Carlo (QMC)
calculations from first-principles. Our theory provides a general
framework to account for the contradictory experimental outcomes.

The paper is organized as it follows.
In Sec.~\ref{improper_symmetry} we present the derivation of the 2D
iron lattice model consistent with the symmetries of the 3D point
group of the FeSe. We show that it is necessary to include improper
rotations and a gauge transformation for translations in order to
define a 2D square lattice model
consistent with the 3D structure. We study what are the
implications of the symmetry constraints for the pairing structure, and show that 
a d$_{xy}$-wave channel is present beside the extended s-wave one.
These predictions are verified by accurate quantum Monte Carlo
simulations, presented in Sec.~\ref{qmc_section}, to determine the
global symmetry of the pairing function from first-principles.
In Sec.~\ref{BCS_modeling} we model the ab-initio pairing function
by means of a BCS Hamiltonian. This allows us to
compute the BCS gap on a dense k-point grid, and study its nodal
structure in the k-space. In Sec.~\ref{physical_consequences} we relate
our findings to the experimental outcome in FeSe, and to other
experiments in different families of iron pnictides and chalcogenides.

\section{Improper $S_4$ symmetry}
\label{improper_symmetry}

\begin{figure}[ht]
\begin{center}
\includegraphics[scale=0.33]{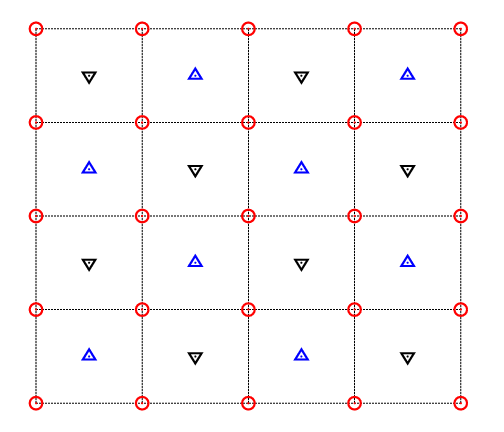}
\caption{Two dimensional square lattice representing the geometry of iron pnictides/chalcogenides
planes in the tetragonal low temperature phase. The iron sites (red
circles) form a perfect square lattice. The Se (As) atoms in e.g. FeSe (BaAs$_2$ Fe$_2$) are below (lower triangles) or above (upper triangles) the iron
layer. It is evident that a symmetry operation like rotation of 90 degrees
around an iron atom, interchanges lower triangles with upper ones. Therefore a
further reflection $\sigma_h$ in the iron plane has to be combined
with  the 90 degree rotation to define a correct symmetry of the
Hamiltonian. In a low energy model where Se and As atomic centers do
not appear, their effect is to define a non trivial spatial symmetry
in the iron square lattice where orbital and spatial degrees of freedom are strongly coupled in an unusual way. As noted in Ref.~\onlinecite{bt}, 
also translation symmetry is not defined, but the composition of translation and reflection $\sigma_h$ 
is again a well defined symmetry of the model. Then, the fictitious broken translation symmetry of the low 
energy model can be easily gauged out (see text and Ref.~\onlinecite{miyake}).  }
\label{fordummies}
\end{center}
\end{figure}

The undistorted structure of FeSe, together with the other parent compounds of the ``111'' and
``1111'' families, belongs to the $P4/nmm$ space-group, which is
nonsymmorphic due to a glide plane parallel to the iron
layers. Indeed, its related $D_{4h}$ point-group classes
contain certain operations with the nonprimitive lattice translation 
$\boldsymbol{\tau} = (1/2,1/2,0)$ - expressed in crystal units of the
2-Fe unit cell - which brings a
Fe site into the other in the cell.
This is due to the Se sites alternating above
and below the Fe layer (see Fig.\ref{fordummies}).

In particular, the $C_4$ subgroup  generated by pure
rotations by $\pi/2$ about the principal axis 
is not 
compatible
with the structure of the solid.
On the other hand, the $S_4$ subgroup, defined with the rotation by
$\pi/2$ about the principal axis combined with the reflection
$\sigma_h$ through the Fe plane ($S_4=C_4 \sigma_h$), is a a pure
point-group operation allowed by the $P4/nmm$ space-group,
as well as the reflection through the plane perpendicular to 
the Fe (xy) plane and bisecting the secondary (x and y) axes (the $\sigma_d$ element).\cite{conventions}

This has important consequences on the symmetries of
the subgroup associated with the 2-dimensional (2D) Fe square
sublattice, and thus on  
the low-energy features of the system, as
for instance, the superconducting  gap structure. 
Indeed, the standard $C_{4\nu}$ group of the 2D square lattice, generated by the elementary $\sigma_d$
and $C_4$ point-group operations, does not fulfill all symmetries
of the $P4/nmm$ space-group, 
whereas
the group
generated by $\sigma_d$ and $S_4$ does.
Moreover, proper ($C_4$) and improper ($S_4$) rotations differ from the physical point of view, 
because a Hamiltonian or a pairing function fulfilling the point-group operations
generated by $S_4$ and $\sigma_d$ will transform in a very different way
with respect to those based on the standard $C_{4\nu}$ group, as it will be shown 
in the following. 

From ab-initio electronic structure DFT and DMFT calculations of
IBS,\cite{singh,aichhorn,werner} it
is well established that the bands crossing the Fermi level have a
strong d-orbital character. Indeed, the minimal tight-binding Hamiltonian which reproduces
successfully the Fermi surface contains all the five atomic (or
Wannier) d-orbitals centered on the Fe sites,\cite{graser} namely $| d_{xz}
\rangle$, $| d_{yz} \rangle$, $| d_{xy} \rangle$, $|d_{x^2-y^2}
\rangle$, and $|d_{3 z^2 -r^2} \rangle$,
expressed with
the crystal axes oriented according to the Fe square
lattice directions.
A given atomic orbital $d_{{\bf R},\nu,\sigma}(\textbf{r})$ defining the 2D multiorbital model can be represented by
$\langle \textbf{r} | d_{{\bf R},\nu,\sigma} \rangle = \langle \textbf{r} |
c^{\dag}_{{\bf R},\nu,\sigma}|0\rangle$, where  ${\bf r}$ lives in the
3D space, ${\bf R}$ 
is the center of the atomic orbital on the square lattice, $\nu$ is the orbital symmetry, 
and $\sigma = \pm \frac{1}{2}$ is the electron spin.
A spatial point-group operation $\eta$
acts on the orbital creation operator by changing its position  
$ {\bf R} \to R_{\eta} {\bf R}$,  its phase $s = 1 \to  s^\prime(\eta,\nu) = \pm 1$ and its orbital
label $\nu \to \nu^\prime(\eta, \nu)$ 
according to the general rule:
\begin{equation}
 R_\eta c^{\dag}_{{\bf R},\nu,\sigma} R^{\dag}_\eta =  s(\eta,\nu)
 c^{\dag}_{R_\eta {\bf R},\nu^\prime (\eta,\nu) ,\sigma}.
\label{general_rules}
\end{equation}
For $\eta=\{\sigma_d,S_4,C_4\}$ and $\nu$ given, the corresponding $s^\prime$ and 
$\nu^\prime$  are reported in Tab.~\ref{sgap}. 
The transformation rules are different, whether the group is generated by $S_4$ or $C_4$.  
\begin{table}[ht]
\begin{ruledtabular}
\begin{tabular}{ l l l l }
 $\nu$  & $\eta=\sigma_d$  & $\eta=S_4$ & $\eta=C_4$  \\
\hline
 $d_{xz}$ & $+d_{yz}$ & $-d_{yz}$ & $+d_{yz}$ \\
 $d_{yz}$ & $+d_{xz}$ & $+d_{xz}$ & $-d_{xz}$ \\
 $d_{xy}$ & $+d_{xy}$ & $-d_{xy}$ & $-d_{xy}$ \\
 $d_{x^2-y^2}$ & $-d_{x^2-y^2}$ & $-d_{x^2-y^2}$ & $-d_{x^2-y^2}$ \\
 $d_{3z^2-r^2}$ & $+d_{3z^2-r^2}$ & $+d_{3z^2-r^2}$ & $+d_{3z^2-r^2}$ \\
\end{tabular}
\caption{ Action of the symmetry operations
  $\eta=\{\sigma_d,S_4,C_4\}$  on the five iron atomic d-orbitals. The
  entries represent the pair $s^\prime=\pm$ and $\nu^\prime=d_{xz},d_{yz},d_{xy},d_{x^2-y^2},d_{3z^2-r^2}$  
as a function of the orbital index $\nu$ (first column) and the symmetry operation 
$\eta$ considered. 
Note the different $s^\prime$ between proper $C_4$ and
improper $S_4$ rotations in the first two orbitals (odd 
with respect to $\sigma_h$).}
\label{sgap}
\end{ruledtabular}
\end{table}

As it is readily seen in Fig.~(\ref{fordummies}), also the elementary
 translation symmetry bringing one iron atom to a nearest neighbor one, 
is not defined on the plane, because, under this transformation, 
 the upper triangles are interchanged with the lower ones.
Again the combination of the conventional translation $T_x$ ($T_y$) 
symmetry with the reflection $\sigma_h$ is 
a well defined  symmetry of the crystal $\tilde T_x = \sigma_h T_x$ ($\tilde T_y = \sigma_h T_y$), 
isomorphic to the conventional translation 
group of the square lattice.
In a low energy model with a given number of orbitals $\tilde T_x$ ($\tilde T_y$) behaves like the 
conventional translation group, only if we take into account a non-translation invariant 
 sign in the definition of 
the localized basis, namely:
\begin{equation} \label{gaugeout}
c^\dag_{{\bf R},\nu,\sigma}  \to (-1)^{x+y} c^{\dag}_{{\bf R},\nu,\sigma}
\end{equation}
for the orbitals $\nu=\{ d_{xz},d_{yz} \}$ that are odd with respect to the symmetry
operator $\sigma_h$. Here ${\bf R}=(xa,ya)$, where $a$ is the lattice space
of the square lattice. Unless otherwise specified, here and
thereafter we use the translation invariant gauge defined in Eq.~\ref{gaugeout}.

With these conventions, let us take into account the 2D tight-binding Hamiltonian
$H_0$ which defines the low-energy model for the IBS. Translation invariance means that all 
the coupling can be written in Fourier space:
\begin{equation} 
{\hat H}_0=\sum_{{\bf k},\nu,\mu,\sigma} t({\bf k})_{\nu,\mu} c^{\dag}_{{\bf
    k},\nu,\sigma}  c_{{\bf k},\mu,\sigma}, 
\label{H_0}
\end{equation}
where the vectors $\{{\bf k}\}$ belong to the unfolded Brillouin zone of the 2D lattice 
with one iron per unit cell,
$c_{{\bf k},\nu,\sigma}$ ($c^\dag_{{\bf k},\nu,\sigma}$) is the Fourier
transform of $c_{{\bf R},\nu,\sigma}$ ($c^\dag_{{\bf R},\nu,\sigma}$),
the $\mu,\nu$ indexes run over the d-orbital labels,
the hoppings fulfill the hermitian condition $t({\bf k})^*_{\mu,\nu} = t({\bf k})_{\nu,\mu}$, 
and $t(-{\bf k})_{\nu,\mu} = t({\bf k})^*_{\nu,\mu}$ holds for reality.
Without using the phase factor $(-1)^{x+y}$ for the odd $\sigma_h$ orbitals,
the hopping between even and odd orbitals would have acquired  an alternating sign, breaking   
the translation symmetry, as correctly pointed out in Ref.~\onlinecite{bt}.
After the simple gauge transformation given in Eq.(\ref{gaugeout}) this fictitious broken symmetry is 
easily removed, and the low energy model becomes translational invariant, with a corresponding 
unfolded BZ.

In our approach we want to  point out that a low energy
two-dimensional model, like the free-band model in Eq.(\ref{H_0}),  
has to be not only translation invariant but has  to satisfy also all point symmetry transformations
generated by $\sigma_d$ and $S_4$, the compatible subgroup of the
full $P4/nmm$ space-group. 
 The latter point group is isomorphic to the standard 
$C_{4\nu}$ group of the square lattice provided, in this group, we identify the 
$\pi/2$ rotation with the corresponding improper one. 

If we apply the transformation rules in Eq.~\ref{general_rules} to
${\hat H}_0$, and
require its invariance, we obtain a set of relations for the hopping matrix:
\begin{equation}
t( R_\eta {\bf k})_{\nu,\mu} = s( \eta,\nu) s(\eta,\mu) t({\bf k})_{\nu^\prime(\eta,\nu),\mu^\prime(\eta,\mu)},
\label{hopping_constraints}
\end{equation}
for the eight possible independent point-group operations, defined by $\eta=\sigma_d$, $S_4$,
and all their products. 
These transformations have already been derived
in Refs.~\onlinecite{nakamura} and \onlinecite{miyake} or implicitly assumed in Refs.~\onlinecite{prx} and 
\onlinecite{graser2}.
However, their implications for the pairing properties have
so far been overlooked.

The pairing operator ${\hat F}$ is assumed to be translationally invariant (on the
square lattice with one iron atom for unit cell and with the sign
convention given in Eq.~\ref{gaugeout} for the odd reflection
orbitals), so that it is convenient to write it in ${\bf k}$-space as 
\begin{equation}
{\hat F}=  \sum_{{\bf k},\nu,\mu} f({\bf k})_{\nu,\mu} 
c^{\dag}_{{\bf k},\nu,\uparrow}  c^{\dag}_{-{\bf k},\mu,\downarrow}.
\label{F}
\end{equation}
Singlet pairing, and no broken time-reversal symmetry imply that
$f({\bf k})_{\nu,\mu} = f(-{\bf k})_{\mu,\nu}$, and
$f({\bf k})_{\nu,\mu} = f(-{\bf k})^*_{\nu,\mu}$,
respectively.
In the following, we derive the spatial-orbital properties of the
pairing function $F({\bf r},{\bf r}^\prime)=\langle {\bf r}, {\bf r}^\prime| {\hat F}
| 0 \rangle$, by assuming
that it belongs to the $A_{1g}$ irreducible representation of the point-group
generated by $\sigma_d$ and $S_4$.
This assumption will be verified later by ab-initio QMC
calculations with a fully optimized Jastrow projected BCS wave function.
``Improper s-wave'' is the name we assign to the symmetry of the
resulting ${\hat F}$, 
to distinguish it from the standard or ``proper'' one, i.e. the $A_{1g}$ irreducible
representation of the isomorphic point-group where the $S_4$ class has
been replaced by $C_4$.

Based on the general rules in Eq.~\ref{general_rules} and the entries of
Tab.~\ref{sgap}, we can derive the symmetry constraints for the
improper s-wave. As the characters of the $A_{1g}$ representation
are - by definition - identically one
for all classes within the point-group, the symmetry transformations
of $f({\bf k})_{\nu,\mu}$ read as
\begin{equation}
f( R_\eta {\bf k})_{\nu,\mu} = s( \eta,\nu) s(\eta,\mu) f({\bf k})_{\nu^\prime(\eta,\nu),\mu^\prime(\eta,\mu)},
\label{improper_s_constraints}
\end{equation}
for $\eta=\sigma_d$, $S_4$, and all their independent products.

The above constraints
reduce the variational freedom for the signs of
$F({\bf r},{\bf r}^\prime)$. Accidental nodes must not break the \emph{general} symmetries in
Eq.~\ref{improper_s_constraints}. For the improper s-wave, we
demonstrate that $F$ has a peculiar, symmetry selected, sign
structure. To show this, let us divide the
iron d-orbitals into two groups, namely the ``even'' ($e$) and``odd''
($o$) ones, according to their parity under $\sigma_h$. 
 The $|d_{xz}\rangle$ and $|d_{yz}\rangle$ orbitals
 are odd, while $|d_{xy}\rangle$, $|d_{x^2-y^2}\rangle$, and $|d_{3
   z^2 -r^2}\rangle$ are even. Thus,
$F$ can be expanded as
\begin{equation}
F= F_{ee}  + F_{eo} + F_{oe}  + F_{oo},  
\label{even_odd}
\end{equation}
with its components defined as follows:
\begin{eqnarray}
F_{ee}({\bf r},{\bf r}^\prime) = & 
\sum_{{\bf k},\{\nu,\mu \} \in \textrm{even}} 
f({\bf k})_{\nu,\mu} 
d_{{\bf k},\nu,\uparrow}({\bf r})  d_{-{\bf k},\mu,\downarrow}({\bf r}^\prime),
\nonumber \\
F_{eo}({\bf r},{\bf r}^\prime) = & 
\sum_{\substack{{\bf k},\{\nu \} \in \textrm{even} \\\{\mu \} \in \textrm{odd} } }
f({\bf k})_{\nu,\mu} 
d_{{\bf k},\nu,\uparrow}({\bf r})  d_{-{\bf k},\mu,\downarrow}({\bf r}^\prime), 
\nonumber \\
F_{oe}({\bf r},{\bf r}^\prime) = & 
\sum_{\substack{{\bf k},\{\nu \} \in \textrm{odd} \\\{\mu \} \in \textrm{even} } }
f({\bf k})_{\nu,\mu} 
d_{{\bf k},\nu,\uparrow}({\bf r})  d_{-{\bf k},\mu,\downarrow}({\bf r}^\prime),
\nonumber \\
F_{oo}({\bf r},{\bf r}^\prime) = & 
\sum_{{\bf k},\{\nu,\mu \} \in \textrm{odd}} 
f({\bf k})_{\nu,\mu} 
d_{{\bf k},\nu,\uparrow}({\bf r})  d_{-{\bf k},\mu,\downarrow}({\bf
  r}^\prime).
\label{F_components}
\end{eqnarray}

The above pairing functions are periodic in both  ${\bf r}$ and ${\bf
  r}^\prime$, with periodicity set by the Bravais lattice vectors.
In the following analysis, without loss of generality, we are going to
pin ${\bf r}$ around an iron site\cite{footnote_iron_center}, while 
we define $F (k_x,k_y)$ as the 2D Fourier transform of $F$ with
respect to the planar ($x,y$)  components 
of ${\bf r}^\prime - {\bf r}$.
According to the transformation rules in
Eq.~\ref{improper_s_constraints} and in Tab.~\ref{sgap}, we obtain that  
$F_{ee}$ and $F_{oo}$ have an s-wave planar symmetry, e.g. 
 $F_{ee} (k_x,k_y) = F_{ee} (-k_x,k_y) = F_{ee} (k_x,-k_y)=F_{ee} (k_y,k_x) $,
whereas 
$F_{eo}$ and $F_{oe}$, which couple
unlike-parity components,  have planar $d_{xy}$ symmetry, e.g. 
$F_{eo} (k_x,k_y) = -F_{eo} (-k_x,k_y) = -F_{eo} (k_x,-k_y)=F_{eo} (k_y,k_x) $.
We note here that in a standard s-wave superconductor obeying the
$C_{4\nu}$ point-group transformations,
also the $F_{eo}$ and $F_{oe}$ components are s-wave. Thus,
our improper-s wave pairing function has to have different signs
just from symmetry considerations.
We note also that a two-band model containing only odd orbitals\cite{prx}, cannot account for 
the  anomalous $d_{xy}$ symmetry of the pairing function. 

\section{Ab-initio Quantum Monte Carlo calculation of the pairing
  function of $FeSe$}
\label{qmc_section}

Our predictions presented in Sec.~\ref{improper_symmetry} and derived from symmetry arguments
have been verified against accurate ab-initio QMC 
calculations based on the energy minimization of a Jastrow
projected antisymmetrized geminal product (JAGP) wavefunction, performed for the FeSe on a $4 \times 4
\times 1$
iron lattice, subject to periodic boundary conditions.
The geminal or pairing function describes the spatial correlations in
the singlet electron pairs, giving rise eventually to a
superconducting state, if the pairing remains stable in the thermodynamic limit.
We used the
experimental FeSe geometry, determined at ambient pressure (0 GPa) and temperature\cite{kumar}
and under a hydrostatic pressure of 4 GPa at low-temperature.\cite{margadonna} Fe and Se atoms have been
replaced by pseudoatoms, containing only 16 and 6
valence electrons, respectively.

\subsection{Variational Jastrow projected AGP wave function} 
\label{variational_wf}

In this work, we used variational quantum Monte Carlo (QMC) methods to find the best many-body wave function $\Psi$ which minimizes the variational energy
\begin{equation}
E = \langle \Psi | H | \Psi\rangle / \langle \Psi | \Psi \rangle,
\label{var_energy}
\end{equation}
where $H$ is the electron first principles Hamiltonian in Hartree units
\begin{equation}
H = -\frac{1}{2} \sum_{i=1}^N \nabla^2_i + \mathop{\sum_{i,j=1}^N}_{i<j} \frac{1}{|{\bf r}_i-{\bf r}_j|} -  \sum_{i=1}^{N_\textrm{nuclei}} \sum_{j=1}^N v_i({\bf R}_i - {\bf r}_j),
\label{H}
\end{equation}
with $N$ the number of electrons in the supercell, $N_\textrm{nuclei}$ the number of nuclei, and ${\bf R}_i$ the position of the i-th  nucleus, fixed at the experimental geometry.
In the Hamiltonian, the core electrons have been replaced by the pseudopotentials $v_i$. For iron, we used the energy-consistent scalar-relativistic Hartree-Fock pseudopotentials generated in Ref.~\onlinecite{burkatzki}, while for selenium, we used the smooth scalar-relativistic Hartree-Fock pseudopotentials published in Ref.~\onlinecite{trail}. The system is unpolarized with $N^\uparrow=N^\downarrow=N/2$, and the supercell with $N$ electrons is subject to periodic boundary conditions
 to reproduce the periodicity of the crystal, with standard Ewald summations of the long range Coulomb potential by including also the constant nucleus-nucleus interaction.

The wave function $\Psi$ is written in the Jastrow projected AGP form
\begin{equation}
\Psi_\textrm{JAGP}({\bf R}_\textrm{el}) = \exp[-J({\bf R}_\textrm{el})] \det[\phi({\bf r}_i^\uparrow,{\bf r}_j^\downarrow)],
\label{JAGP}
\end{equation}
where $1 \le i,j \le N/2$, and ${\bf R}_\textrm{el}=\{ {\bf r}_1^\uparrow,\ldots,
{\bf r}_{N/2}^\uparrow, {\bf r}_1^\downarrow,\ldots,
{\bf r}_{N/2}^\downarrow\}$  the many-body electron configuration.
The variational form above has been widely used to study strongly
correlated Hamiltonians on the lattice\cite{tj0,mott,tj1}, and then
generalized to a large variety of ab-initio systems, ranging from
molecules\cite{casula_mol,sandro_benzene,spanu_water} to
solids\cite{sandro_silicon,mariapia_graphene}. In all systems studied
so far, it has been proven to be accurate and capable to describe
strong charge localization, polarization effects and weak dispersive
forces. The functional form in Eq.~\ref{JAGP} is very flexible. The
Jastrow factor $J$ reads as 
\begin{equation}
J (\textbf{r}_1, \cdots , \textbf{r}_N) = \sum_{i=1}^{N_\textrm{nuclei}} \sum_{j=1}^N g^\textrm{1-body}_i({\bf R}_i - {\bf r}_j) +  \mathop{\sum_{i,j=1}^N}_{i<j} g( \textbf{r}_i, \textbf{r}_j ),
\label{J}
\end{equation}
where $g^\textrm{1-body}_i$ is the electron-nucleus term, while $g$ includes electron-electron correlations. We used a spin-independent parametrization for $J$, therefore the spin indexes are dropped out from Eq.~\ref{J}. The one-body part is developed on Gaussian orbitals $\chi_l^i$ (with $l$ basis set index, and $i$ nuclear index), and it depends on the i-th nucleus as follows
\begin{equation}
g^\textrm{1-body}_i({\bf R}_i - {\bf r})=\sum_l G_l^i ~ \chi_l^i(\vc{R}_i-\vc{r}),
\end{equation}
where the Gaussian basis set is built of $2s2p/[1s1p]$ contracted  $\chi_l^i$ orbitals for selenium, while we used a $2s2p2d/[1s1p1d]$ contracted Gaussian basis set for iron. Both $ G_l^i$, the Gaussian exponents, and the linear coefficients of the contractions are variational parameters to be optimized.
Note that there are no electron-nucleus cusps, as the pseudopotentials in Refs.~\onlinecite{burkatzki} and \onlinecite{trail} are finite at the origin.
The electron-electron part $g$ of the Jastrow factor is defined as
\begin{equation}
g( \textbf{r}, \textbf{r}^\prime)=u(|\textbf{r}-\textbf{r}^\prime|) + \sum_{ijlm} H_{lm}^{ij} ~ \chi_l^i(\vc{R}_i-\vc{r}) \chi_m^j(\vc{R}_j-\vc{r}^\prime),
\end{equation}
where the homogeneous part $u(r)=0.5 r/(1+\alpha r)$ fulfills the electron-electron cusp conditions for unlike-spin particles, and the Gaussian basis sets $\chi_l^i$ are the same as the ones in the one-body part.  $\alpha$ and the symmetric matrix $H_{lm}^{ij}$ are variational parameters. The basis set parameters (Gaussian exponents and the linear coefficients) are optimized together with the one-body part.

The pairing function $\phi$ in Eq.~\ref{JAGP} reads as
\begin{equation}
\phi({\bf r},{\bf r}^\prime)= \sum_{i=1}^M \lambda_i \psi_i^{MO}({\bf
  r}) \psi_i^{MO}({\bf r}^\prime), 
\label{pairing_MO}
\end{equation}
where the sum is over $M (\ge N/2)$ molecular orbitals $\psi_i^{MO}(\textbf{r})$, where each one 
can be occupied by opposite spin electrons. The orbitals $\psi_i^{MO}(\textbf{r})$ are expanded in a Gaussian single-particle basis set $\{\chi^\textrm{det}_j\}$, centered on the atomic nuclei as in the Jastrow 
case, i.e. 
\begin{equation}
\psi_i^{MO}(\textbf{r})=\sum_j K_{ij}\chi^\textrm{det}_j (\textbf{r}),
\end{equation}
where the sum in the above equation runs over both the basis set and nuclear center indexes 
(for simplicity of notations we omit here the explicit dependence on the atomic positions). 
The $\{\chi^\textrm{det}_{i}\}$ basis set is uncontracted with size of $9s8p7d6f$ for Fe and $6s5p4d3f$ for Se. $\lambda_i$, $K_{ij}$, and the exponents of the uncontracted Gaussian basis set $\{\chi^\textrm{det}_j\}$ are variational parameters.
If $M=N/2$, the expansion in Eq.~\ref{pairing_MO} is equivalent to a single Slater determinant, when the antisymmetrization of $\phi$ is performed. On the other hand, if  $M > N/2$, correlation is introduced through the pairing function, and one can describe the effective attraction between opposite spin electrons by a reduction of the total variational energy. If this energy gain survives in the thermodynamic limit, it implies that the BCS condensate formed by the electron singlets is stable, and the symmetry of the pairing function is the one of the superconducting gap. 

In order to treat the periodic system in a finite supercell, both the Jastrow Gaussian basis set $\{\chi_l^i\}$ and the one $\{\chi^\textrm{det}_j\}$ of the pairing function $\phi$ are made of periodic Gaussian functions, as described in Ref.~\onlinecite{sandro_silicon}.

\subsection{Numerical methods}
\label{numerical_methods}

To generate the initial molecular orbitals, we performed preliminary density functional theory (DFT) calculations in the local density approximation, within the same basis described in the previous subsection. The starting Gaussian exponents of the $\{\chi^\textrm{det}_j\}$ basis are even-tempered, as they are defined as $ Z_i = \alpha \beta^{i}$ for $i=0,\ldots,n-1$ with $\alpha=Z_\textrm{min}$ and $\alpha \beta^{n} = Z_\textrm{max}$.
The number of Gaussians $n$, $Z_\textrm{min}$ and $Z_\textrm{max}$ are optimized independently for each angular momentum shell in order to minimize the total DFT energy. The number of shells, and the number of Gaussians per shell are chosen to give a convergence of $\approx$ 10 mHartree per iron atom in the DFT energy.\cite{azadi}

The MO generated by DFT are then used to build the $\Psi_\textrm{JAGP}$ wave function. At the beginning, the pairing function is such that $\lambda_i =1$ for $ i\le N/2$ and $\lambda_i =0 $ for $ i > N/2$, i.e. we start from a Jastrow-Slater wave function, with the Slater part exactly given by the DFT orbitals and with no Jastrow factor ($H^{ij}_{lm}=G^i_l=0$) apart from an initial guess of the variational constant $\alpha \approx 1$ in its homogeneous part $u$. 
The many-body wave function is optimized in the QMC framework by minimizing the variational energy in Eq.~\ref{var_energy}, 
by using the stochastic reconfiguration (SR)  method (based on the first-order derivatives of the energy with respect to the variational parameters) combined with second-order contributions coming from the Hessian matrix (``linear method''). Indeed, the SR method described in Ref.~\onlinecite{casula_mol} has been improved by using a statistically more accurate
evaluation of the Hessian matrix elements in the so called ``linear basis'' as described in Ref.~\onlinecite{sorella_hessian}.
Before the full optimization of all the variational parameters in our 
JAGP wave function, we adopt  the following intermediate steps:  
\begin{description}
\item{i)} first the Jastrow linear coefficients $ G_l^i$ and $H_{lm}^{ij}$, together with the parameter $\alpha$ of the homogeneous pairwise Jastrow factor, 
are optimized;
\item{ii)} then, also the linear coefficients of the Jastrow basis contractions $\{\chi_l^i\}$  are
 optimized; 
\item{iii)} not only the contraction coefficients but also the Jastrow Gaussian exponents are relaxed, together with all previous parameters, by means of the linear minimization method;
\item{iv)} finally, all variational parameters (in both the Jastrow and the Slater determinant) are optimized. In this last step the molecular orbitals evolve according  
to the algorithm described in Ref.\onlinecite{mariapia_molecules}, that allows one to exploit the derivatives with respect to their atomic (uncontracted) components without changing the rank of the full pairing matrix.
In the last step, in order to decrease the number of variational parameters involved 
in the uncontracted basis for the AGP, we perform the minimization by using only one contracted orbital for 
each angular momentum. This allows us to represent in the most efficient way 
the correction to the DFT orbitals.
With this strategy we achieve a dramatic reduction of the number of parameters and a systematic improvement of the variational 
energy, which substantially decreases with respect to step (iii). 
\end{description}

Once the best Jastrow correlated single determinant wave function is
obtained, we let the wave function develop a singlet pairing
correlation, by expanding it beyond the single Slater determinant. We
add in Eq.~\ref{pairing_MO} the first $M-N/2$ additional DFT MO
orbitals above the Fermi level, with their corresponding 
$\lambda_i$ (with $i$ s.t. $N/2< i \le M$)
taken from an educated guess based on a BCS-like estimate of the
pairing amplitude, and later fully optimized. 
The optimal $M$ is chosen to match the criterion discussed in Ref.~\onlinecite{mariapia_molecules}, namely it is the sum of the majority spin orbitals in the atomic limit ($M = \sum_{i=1}^{N_\textrm{nuclei}} n_i^\uparrow$, where we take the convention that $n_i^\uparrow \ge n_i^\downarrow ~~ \forall i$ in the dissociation limit).
The final global optimization is performed with the same method used before in (iv), with all parameters free to move, and the convergence of the $\Psi_\textrm{JAGP}$ is reached if the variation of $\lambda_i$ with $i > N/2$ is within the statistical noise. Note that the hierarchy followed in the wave function optimization is related to the  energy scale of its various components, the smallest and the last one being the energy gain (also called ``condensation energy'') due to the paired state with respect to the Fermi liquid reference, namely 
$\lambda_i=0$ for $i> N/2$. 

The final wave function yields invaluable information about the correlated ground state, like the symmetry of the superconducting gap, in the case that a paired state is stabilized by the variational approach. This method is unbiased, as we do not provide any constraint in the energy minimization, except for the symmetries imposed by the translational invariance of the lattice structure, which reduce the total number of independent variational parameters. In a large variety of systems studied by this wave function, we reached the chemical accuracy (in the energy differences), without resorting to further projective methods like diffusion Monte Carlo. We would like also to highlight the fact that this variational method is sign problem free.

For the largest system computed, the FeSe $4 \times 4 \times 1$
lattice (16 Fe + 16 Se) with 352 electrons, a full QMC energy
minimization has been performed to optimize a total amount of about
10000 independent variational parameters. 

\subsection{Results}
\label{qmc_results}

The full many-body wave
function is written in a compact form as in Eq.~\ref{JAGP}.
In all the calculations quantum effects on nuclei are neglected, within the Born-Oppenheimer 
approximation. Hence superconductivity is assumed to be non conventional, namely 
not coming from electron-phonon coupling.

\begin{figure}[ht]
\includegraphics[width=\columnwidth]{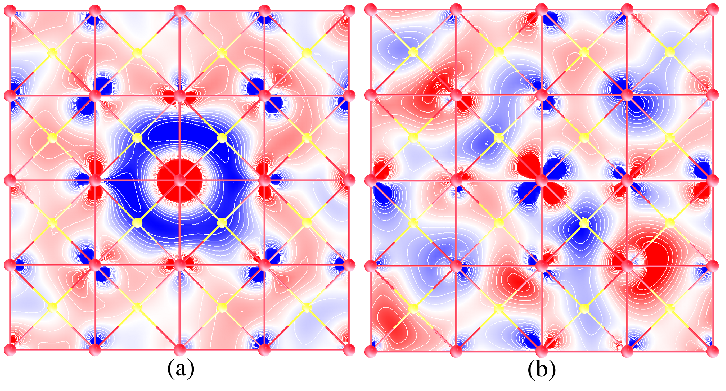}
\caption{
Even-even (panel a), and odd-even (panel b) components of the pairing
function $\phi^{corr}$ of the FeSe high-T$_c$ superconductor at $0$ GPa
obtained by QMC, by optimizing the energy of the variational 
wavefunction without assuming any 
point symmetry.
The contour plots show $\phi^{corr}({\bf R}_\textrm{center},{\bf r})$ with
${\bf R}_\textrm{center}$ set to be the iron lattice site at the center of the
supercell, while ${\bf r}$  
spans the plane defined by the $4 \times 4$ lattice.
Red (yellow) balls are iron (selenium) sites.
Arbitrary units blue (red) intensity indicates negative (positive)  regions with corresponding magnitude. 
}
\label{pairing_0GPa}
\end{figure}

\begin{figure}[ht]
\includegraphics[width=\columnwidth]{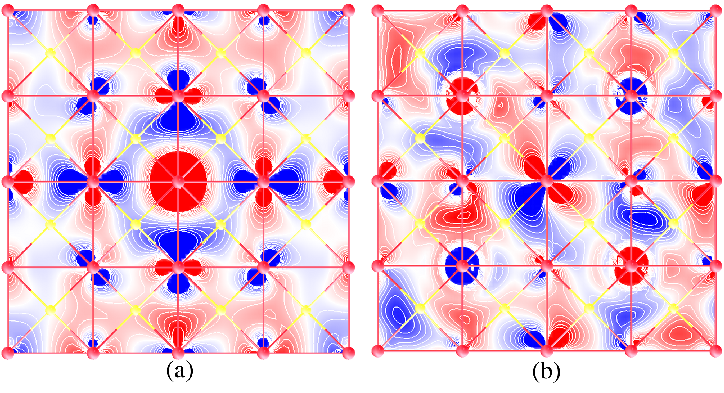}
\caption{The same as in Fig.~\ref{pairing_0GPa} but for a structure
  corresponding to a pressure of $4$ GPa.
}
\label{pairing_4GPa}
\end{figure}

$\phi({\bf r},{\bf r}^\prime)$, defining the wave function in Eq.~(\ref{JAGP}), 
is an extension of $F({\bf r},{\bf
  r}^\prime)$ in Eq.~\ref{even_odd}, as it is expanded over the full
atomic basis set used to represent the MO's. 
The resulting $\phi$,
restricted to its correlated part $\phi^{corr}$ (obtained by summing
the MOs in Eq.~\ref{pairing_MO} over $i$ with $N/2 < i \le M$), is
plotted in Figs.~\ref{pairing_0GPa} and 
~\ref{pairing_4GPa} for 
0 and 4 GPa, respectively.
$\phi^{corr}$ has been projected
over its 4 possible components $\phi^{corr}_{ee}$
(Figs.~\ref{pairing_0GPa}(a) and \ref{pairing_4GPa}(a)),
$\phi^{corr}_{oo}$, $\phi^{corr}_{eo}$, and 
$\phi^{corr}_{oe}$ (Figs.~\ref{pairing_0GPa}(b) and
\ref{pairing_4GPa}(b)), based on the parity with respect to
$\sigma_h$ 
\cite{note_odd}. 

The final picture
 is consistent with the improper s-wave pairing
 function 
predicted by
the 2D point-group symmetries and under the
assumption of a 3D $A_{1g}$ irreducible representation. 
Our QMC solution is $A_{1g}$ as a
 consequence of an unbiased energy minimization, and it has not been
 imposed a priori in the optimization. 
Thanks to the improper $S_4$ generated point-group, the pairing
function $\phi$ shows two planar
 symmetry channels, s-wave for the $\phi^{corr}_{ee}$ and
 $\phi^{corr}_{oo}$, d$_{xy}$-wave for the $\phi^{corr}_{eo}$ and
$\phi^{corr}_{oe}$ components.
We emphasize that 
$\phi$
does not possess 
a well defined symmetry under a proper rotation by $\pi/2$, just because
such a rotation is not a symmetry of the crystal. 
Moreover, Fig.~\ref{pairing_4GPa} shows that this picture
holds even when the tetragonal point-group symmetry is slightly broken, 
and lowered to the orthorhombic point-group, as in the case of applied
pressure and/or low temperature.

\begin{figure}[ht]
\includegraphics[width=\columnwidth]{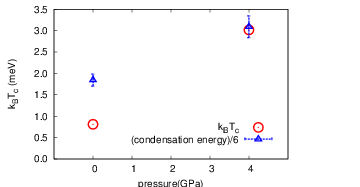}
\caption{Condensation energy in the $4 \times 4$ system (in Hartree), and pairing gap
  $\Delta_{BCS}$ (in meV)  at $0$
  and $4$ GPa, obtained from our QMC calculations, compared
  to the experimental superconducting temperature T$_c$ in meV.
}
\label{condensation}
\end{figure}

To further check the reliability of $\Psi_\textrm{JAGP}$
to describe the superconducting features of FeSe, we
estimated (via correlated sampling) the condensation 
energy $\epsilon_{cond}$,
defined as the difference (per iron atom) between
the expectation value of $H$ computed
with the fully paired $\Psi_\textrm{JAGP}$
and the one with the Jastrow projected single determinant,
obtained by summing the MOs with
$ i \le N/2$ in  Eq.~\ref{pairing_MO}.
Those quantities are much more
sensitive to finite size errors than the pairing symmetry, set by
the short-range 
behavior
of $\phi$, which is converged in the $4 \times 4$
supercell. 
The $\epsilon_{cond}$ 
is  shown in Fig.~\ref{condensation}, together with
the experimental critical temperature $T_c$.
It turns
out that  $\epsilon_{cond} \approx  6 k_B T_c$, which is a remarkable result
given the tiny energy scale (of the order of 1 meV) involved in the gap opening.\cite{foot_temperature}
Moreover, it follows quite well the
behavior of the experimental $T_c$, which increases with pressure,
in the pressure range considered here\cite{foot_temperature}. 
Therefore, if we assume that the condensation energy 
should be related to the critical temperature, as it is a common feature of 
high-$T_c$ superconductors, our pairing function describes reasonably 
well the experiments.
Remarkably, 
we are close to make 
quantitative predictions on unconventional high temperature 
superconductors by means of a fully ab-initio method.

\section{BCS modeling of the QMC pairing function}
\label{BCS_modeling}
The determinantal part of our JAGP variational wave function, for technical reasons,  has not a simple interpretation in terms of a mean-field 
BCS Hamiltonian, as it is common practice in lattice model calculations 
of strongly correlated systems\cite{spanu,tj1}.
In this case, it is well known that very few short 
range parameters can describe very accurately the gap function
$\Delta({\bf k})$ of 
a correlated superconductors.
Moreover, whenever the variational description is accurate enough, very small 
supercells are necessary to describe non trivial physical effects and also 
phase transitions\cite{3chains}. 
For the above reasons, 
it is therefore extremely important to describe our variational wave function
in terms of a mean-field BCS picture, because it helps to reduce the finite size 
effects.  Otherwise, a brute force variational approach based on the JAGP, would require 
high resolution in momentum space, and therefore too large supercells,  
to describe correctly the opening of the 
superconducting gap over the non trivial and rather involved hole and electron 
Fermi surfaces of pnictides. 
For example, we will show in the following Subsections, that a short range parametrization of an improper s-wave gap function  
represents in a quantitative way our paired variational  state.
This is extremely useful to obtain a  detailed information 
in ${\bf k}-$space, and finally  provides 
 a rather satisfactory description of the 
realistic material.

\subsection{BCS model Hamiltonian and pairing function}
\label{bcs_equations}

We start with the five band model defined on the iron plane in real space:
\begin{eqnarray}
H_{BCS} &=& \sum\limits_{ij,\mu,\nu,\sigma} t({\bf R}_i-{\bf R}_j)_{\mu,\nu}
c^{\dag}_{{\bf R}_i,\mu,\sigma} c_{{\bf R}_j,\nu,\sigma} \nonumber \\
&+& \sum\limits_{ij,\mu,\nu} \Delta({\bf R}_i-{\bf R}_j)_{\mu,\nu} c^{\dag}_{{\bf
    R}_i,\mu,\uparrow} c^{\dag}_{{\bf R}_j,\nu,\downarrow}+ {\rm h.c.} 
\label{BCS_model_real_space}
\end{eqnarray}
We assume no broken time reversal symmetry and therefore all couplings are real, including the BCS ones, and singlet pairing.
Thus both $t$ and $\Delta$ have to be considered symmetric real matrices.

Due to translation invariance the model is more easily written in
${\bf k}-$space:
\begin{eqnarray}
H_{BCS} &=& \sum\limits_{k,\mu,\nu,\sigma} t({\bf k})_{\mu,\nu}
c^{\dag}_{{\bf k},\mu,\sigma} c_{{\bf k},\nu,\sigma} \nonumber \\
&+& \sum\limits_{k,\mu,\nu} \Delta({\bf k})_{\mu,\nu} c^{\dag}_{{\bf
    k},\mu,\uparrow} c^{\dag}_{-{\bf k},\nu,\downarrow}+ {\rm h.c.} 
\label{BCS_model_k_space}
\end{eqnarray}
It is easy to show that, for each ${\bf k}$, both matrices $t$ and $\Delta$ may be complex, but they have to be hermitian.
In {\bf k}-space the relations in Tab.(\ref{sgap}) imply parity symmetry constraints 
on the $t({\bf k})$ matrix elements, as we have seen in
Sec.~\ref{improper_symmetry}. The same constraints hold for the BCS
pairing  
matrix $\Delta({\bf k})$. In case of a "generalized" 
$A_{1g}$ real pairing (dubbed here "improper s-wave") the symmetry
relations are the following:
\begin{widetext}
\begin{align}
\Delta_{\mu,\mu}(k_x,k_y)&=&\Delta_{\mu,\mu}(\pm k_x,\pm k_y)
&=&\Delta_{\mu,\mu}(k_y,k_x)  & &\textrm{for  } \mu\ne xz,yz  \nonumber \\
\Delta_{\mu,\mu}(k_x,k_y)&=&\Delta_{\mu,\mu}(\pm k_x,\pm k_y)  &&  &&
\textrm{for  } \mu=xz,yz \nonumber\\
\Delta_{xz,xz}(k_x,k_y)    &=&\Delta_{yz,yz}(k_y,k_x)    && && \nonumber \\ 
\Delta_{xz,yz}(k_x,k_y)  &=&-\Delta_{xz,yz}(-k_x,k_y) & =& -\Delta_{xz,yz}(k_x,-k_y) &=&\Delta_{yz,xz} ( k_y,k_x)  \nonumber \\
\Delta_{xz,x^2-y^2}(k_x,k_y)&=&\Delta_{xz,x^2-y^2}(-k_x,k_y)&=&-\Delta_{xz,x^2-y^2}(k_x,-k_y)&&\nonumber \\
\Delta_{yz,x^2-y^2}(k_x,k_y)&=&-\Delta_{yz,x^2-y^2}(-k_x,k_y)&=&\Delta_{yz,x^2-y^2}(k_x,-k_y)
&=&-\Delta_{xz,x^2-y^2}(k_y,k_x) \nonumber \\
\Delta_{xz,xy}(k_x,k_y)&=&-\Delta_{xz,xy}(-k_x,k_y)&=&\Delta_{xz,xy}(k_x,-k_y)&& \nonumber \\
\Delta_{yz,xy}(k_x,k_y)&=&\Delta_{yz,xy}(-k_x,k_y)&=&-\Delta_{yz,xy}(k_x,-k_y) &=&\Delta_{xz,xy}(k_y,k_x)\nonumber \\
\Delta_{xz,z^2}(k_x,k_y)&=&\Delta_{xz,z^2}(-k_x,k_y)&=&-\Delta_{xz,z^2}(k_x,-k_y)&&\nonumber \\
\Delta_{yz,z^2}(k_x,k_y)&=&-\Delta_{yz,z^2}(-k_x,k_y)&=&\Delta_{yz,z^2}(k_x,-k_y) &=&\Delta_{xz,z^2}(k_y,k_x)\nonumber \\
\Delta_{x^2-y^2,xy}(k_x,k_y)&=&-\Delta_{x^2-y^2,xy}(-k_x,k_y)&=&-\Delta_{x^2-y^2,xy}(k_x,-k_y) &=&-\Delta_{x^2-y^2,xy} ( k_y,k_x)\nonumber \\
\Delta_{x^2-y^2,z^2}(k_x,k_y)&=&\Delta_{x^2-y^2,z^2}(-k_x,k_y)&=&\Delta_{x^2-y^2,z^2}(k_x,-k_y) &=&-\Delta_{x^2-y^2,z^2} ( k_y,k_x)\nonumber \\
\Delta_{xy,z^2}(k_x,k_y)&=&-\Delta_{xy,z^2}(-k_x,k_y)&=&-\Delta_{xy,z^2}(k_x,-k_y) &=&\Delta_{xy,z^2} ( k_y,k_x)\nonumber \\
\label{delta_symmetry_relations}
\end{align}
\end{widetext}
Notice that, by the no broken time reversal condition (real space
couplings, i.e. $t({\bf k}) = t^*(-{\bf k})$,$\Delta({\bf k})=\Delta^*(-{\bf k})$), 
the above functions are purely real or purely imaginary depending on the 
transformation under reflection, even ($\Delta_{\mu,\nu} ({\bf k})=+
\Delta_{\mu,\nu} (-{\bf k})$) or odd ($\Delta_{\mu,\nu} ({\bf
  k})=-\Delta_{\mu,\nu} (-{\bf k})$), respectively.

A generic quasiparticle with spin up  can be written as:
\begin{equation}
\psi^{\dag}_{{\bf k},\nu,\uparrow} = \sum\limits_{\mu} u_{\mu,\nu} c^{\dag}_{{\bf k},\mu,\uparrow}  + v_{\mu,\nu} c_{-{\bf k},\mu,\downarrow}
\label{qp_wf}
\end{equation} 
where $u$ and $v$ are arbitrary complex $5\times 5$ matrices, satisfying:
\begin{eqnarray}
 u^\dag u + v^\dag v &=& I \nonumber \\
 u^\dag v - v^\dag u &=& 0  
\label{nonso}
\end{eqnarray} 
In order to compute these matrices it is enough to write the quasiparticle equation:
\begin{equation}
 \left[ H_{BCS}, \psi^{\dag}_{{\bf k},\nu,\uparrow} \right] = E_\nu({\bf k}) \psi^{\dag}_{{\bf k},\nu,\uparrow} 
\label{qp_equation}
\end{equation}
which implies to solve a $10 \times 10$ eigenvalue problem for each ${\bf k}$:
\begin{equation} 
\label{bcseq}
\begin{array}{|cc||c|}
 t({\bf k})  & \Delta({\bf k}) & u_\nu \\
 \Delta({\bf k}) & -t({\bf k})  & v_\nu 
\end{array} = E_\nu({\bf k})
\begin{array}{|c|}
 u_{\nu} \\
 v_{\nu} 
\end{array}
\end{equation}
Notice that 
Eq.~\ref{bcseq} 
has pairs of eigenvalues of opposite values as if $(u,v)$ is an eigenvector 
with eigenvalue $E({\bf k})$, $(v,-u)$ has eigenvalue $-E({\bf k})$. Thus we have 5 positive and 
5 negative eigenvalues for each {\bf k}.
The pairing function $f^{BCS}$ defines the ground state $\Psi_{BCS}$ 
 of the BCS Hamiltonian by:
\begin{equation}
|\Psi_{BCS} \rangle = \exp \left(\sum\limits_{\mu,\nu}  f^{BCS} ({\bf k})_{\mu,\nu} c^{\dag}_{{\bf k},\mu,\uparrow} 
c^{\dag}_{-{\bf k},\nu,\downarrow} \right) | 0 \rangle 
\label{bcs_wf}
\end{equation} 
The function $f^{BCS}$ can be determined by requiring that the ground state is the 
vacuum of all the quasiparticles with positive eigenvalues:
\begin{equation}
\psi_{{\bf k},\nu,\uparrow} | \Psi_{BCS} \rangle =0
\end{equation} 
for all eigenvalues $E_\nu({\bf k}) > 0$.
By using that
\begin{widetext}
\begin{equation}
\exp( -F^\dag ) \psi_{{\bf k},\nu,\uparrow} \exp( F^\dag ) = 
\sum\limits_{\mu} \left\{ u^\dag_{\nu,\mu} c_{{\bf k},\mu,\uparrow} 
+\left[(u^\dag f^{BCS})_{\nu,\mu} + v^\dag_{\nu,\mu} \right] c^\dag_{-{\bf k},\mu,\downarrow} \right\},
\end{equation}
\end{widetext}
where $ F^\dag= \sum\limits_{\mu,\nu}  f^{BCS}({\bf k})_{\mu,\nu} c^{\dag}_{{\bf k},\mu,\uparrow} 
c^{\dag}_{-{\bf k},\nu,\downarrow}$,
we finally  obtain that the matrix $f^{BCS}$ is fully determined via
the quasiparticle equation.  By solving $u^\dag f^{BCS} +v^\dag =0$, it reads:
\begin{equation}
 f^{BCS}({\bf k}) =-(u^\dag)^{-1} v^{\dag},
\label{f_solution}
\end{equation}
namely, for each ${\bf k}$, $f^{BCS}$ is a generally full $5\times 5$ matrix. 
From Eq.(\ref{nonso}) it also follows that $f^{BCS}({\bf k})$ is hermitian for each ${\bf k}$ and can 
therefore be diagonalized with real eigenvalues.
This property is the final consequence of the assumed no broken
time reversal symmetry, otherwise $f^{BCS}$ is only a symmetric matrix, 
generally complex.

\subsection{BCS fitting of the QMC pairing function}
\label{k-resolved_coupling}

We use the BCS model of Eq.~\ref{BCS_model_real_space} as a way to
interpolate on an ultra dense k-grid the realistic pairing 
$f$ which enters the JAGP wave function, computed on a $4 \times 4$
k-point mesh by the ab-initio variational QMC method as described in
Sec.~\ref{qmc_section}.
Within the BCS framework, we
fit both the tight-binding model and the effective BCS coupling
$\Delta$ from ab-initio calculations. The tight-binding model in
Eq.~\ref{BCS_model_k_space} is fitted based on the Wannier
interpolation of the LDA-DFT band structure\cite{footnote_pwscf}
calculated for the undistorted ambient pressure FeSe compound. The
number of neighbors taken in the model is the same as the one 
used in Ref.~\onlinecite{graser} by Graser \emph{et al.}. 
On the other hand, we assume a short-range BCS $\Delta$, by taking
only the nearest-neighbors couplings, assumption that is normally
verified for strongly correlated superconductors. Those couplings are
set to reproduce the real space image of the ab-initio QMC pairing
function $\phi^{corr}({\bf r},{\bf r}^\prime)$ plotted in
Fig.~\ref{pairing_0GPa}. 
 We define the real-space BCS pairing function as:
\begin{equation}
F^{BCS}({\bf r},{\bf r}^\prime) = \sum_{{\bf k},\nu,\mu} 
f^{BCS}({\bf k})_{\nu,\mu} d^G_{{\bf k},\nu,\uparrow}({\bf r})  d^G_{-{\bf k},\mu,\downarrow}({\bf r}^\prime),
\end{equation}
where $f^{BCS}({\bf k})_{\nu,\mu}$ is the solution of the BCS problem (\ref{BCS_model_k_space})
on a given k-grid as in Eq.~\ref{f_solution}, and 
$d^G_{{\bf k},\nu,\uparrow}({\bf r})$ are localized
functions taken as Gaussian periodic orbitals with d-symmetry\cite{sandro_silicon} and
Gaussian exponent $Z=0.8$. 
Moreover, according to Eq.~(\ref{gaugeout}), an overall sign $(-1)^{x+y}$ is given to the 
localized Wannier orbitals, centered at ${\bf R}=(ax,ay)$ and  with odd reflection symmetry. 
This essentially implies that 
$d^G_{{\bf k},\nu,\uparrow}({\bf r}) \to d^G_{{\bf
    k+Q},\nu,\uparrow}({\bf r})$, for such odd reflection orbitals,
where ${\bf Q}=(\pi,\pi)$.  In this way the anomalous part 
of the pairing function $F_{eo}+F_{oe}$ 
that connects even and odd $\sigma_h$ orbitals, breaks the translation 
symmetry because couples ${\bf k}$ and ${\bf k}+{\bf Q}$ physical momenta.
However, this broken translation symmetry, noticed in
Refs.~\onlinecite{bt} and \onlinecite{miyake}, 
 can be completely gauged out in the low energy 2D model,
and is not so crucial therefore (e.g. absolutely irrelevant for the spectrum), 
as the improper $s-$wave symmetry we are pointing out in this work.

 We choose the 
non-zero couplings $\Delta$ in Eq.~\ref{BCS_model_real_space} such
that $F^{BCS}({\bf r},{\bf r}^\prime)$ is as close as possible to the
QMC $\phi^{corr}({\bf r},{\bf r}^\prime)$ in the largest QMC
supercell, by minimizing the distance
in the $L2$ functional space. This is obtained by maximizing the
normalized scalar product $F^{BCS} \cdot \phi^{corr}/\sqrt{\| F^{BCS} \|~ \|
    \phi^{corr}\|}$, which in the Gaussian basis reads:
\begin{equation}
\frac{{\rm Tr}[ S f^{BCS} S
  \tilde{\phi}^{corr} ]}{\sqrt{{\rm Tr}[ S f^{BCS} S f^{BCS}]~ {\rm Tr}[ S
    \tilde{\phi}^{corr} S  \tilde{\phi}^{corr} ]}},
\end{equation}
where $\tilde{\phi}^{corr}$ is the projection of $\phi^{corr}$ on the
single Gaussian basis set $d^G_{{\bf R},\nu}({\bf r})$, and
$S=S({\bf R})_{\mu \nu} = \langle d^G_{{\bf 0},\mu} | d^G_{{\bf
    R},\nu} \rangle$ is the overlap matrix between the basis set
elements.
The largest normalized scalar product found is about $50 \%$, with best
couplings which turn out to be (in eV):
\begin{align}
\Delta_{xz,xz}(\delta {\bf R}_x) &=& \Delta_{yz,yz}(\delta {\bf R}_y) &=&
-0.080 \nonumber \\
\Delta_{xz,xz}(\delta {\bf R}_y) &=& \Delta_{yz,yz}(\delta {\bf R}_x) &=&
-0.051 \nonumber \\
\Delta_{xy,xy}(\delta {\bf R}_x) &=& \Delta_{xy,xy}(\delta {\bf R}_y) &=&
0.158 \nonumber \\
\Delta_{z^2,z^2}(\delta {\bf R}_x) &=& \Delta_{z^2,z^2}(\delta {\bf R}_y) &=&
-0.0163 \nonumber \\
\Delta_{x^2-y^2,z^2}(\delta {\bf R}_x) &=& -\Delta_{x^2-y^2,z^2}(\delta {\bf R}_y) &=&
0.0035 \nonumber \\
\Delta_{xz,xy}(\delta {\bf R}_x) &=& \Delta_{yz,xy}(\delta {\bf R}_y) &=&
-0.130,
\label{delta_bcs_couplings}
\end{align}
where $\delta {\bf R}_x = (\delta,0)$, $\delta {\bf R}_y =
(0,\delta)$, with $\delta$ the distance between two nearest-neighbors
iron sites. The last one in Eq.~\ref{delta_bcs_couplings} is the
even-odd/odd-even coupling which triggers the planar d-wave component and
leads to the improper s-wave symmetry of the global pairing.

It is interesting to note that the resulting
pairing $f({\bf k})_{\mu,\nu}$ is purely real if it couples
even-even and odd-odd orbitals, while it is purely imaginary when it
couples even-odd and odd-even orbitals. This follows from the
$t({\bf k})$ and $\delta({\bf k})$ components, which are either purely
real or purely imaginary according to their transformation properties,
as already highlighted in the previous Subsection.

In the literature, the k-resolved image of the coupling and its symmetry analysis provided
so far, are usually done for the $\tilde{f}({\bf k})_{m,n}$ pairing function
projected and rotated in the band space (here $m$ and $n$ refer to
band indexes), i.e. $\tilde{f}({\bf k})=U^\dagger({\bf k}) f({\bf k})
U({\bf k})$, with $U$ the unitary matrix mapping the orbital into the
band space. We stress that the band space is not the ideal basis to
study the improper s-wave symmetry, as only the orbitals are
eigenstates of the parity under reflection through the iron plane.
The hybrid nature of the bands does not allow to separate the planar s- and
d-wave components of the total paring operator ${\hat F}$ in
Eq.~\ref{F}. In the next section, we will analyze in detail the
k-resolved components of $F$ in the orbital basis and the
superconductivity gap of the associated BCS model.  Here, we provide
the k-resolved image of $\tilde{f}({\bf k})_{m,n}$ in the band
basis on the Fermi surface, in order to understand its sign
properties, and compare it with the previous literature. 

\begin{figure}[ht]
\includegraphics[width=\columnwidth]{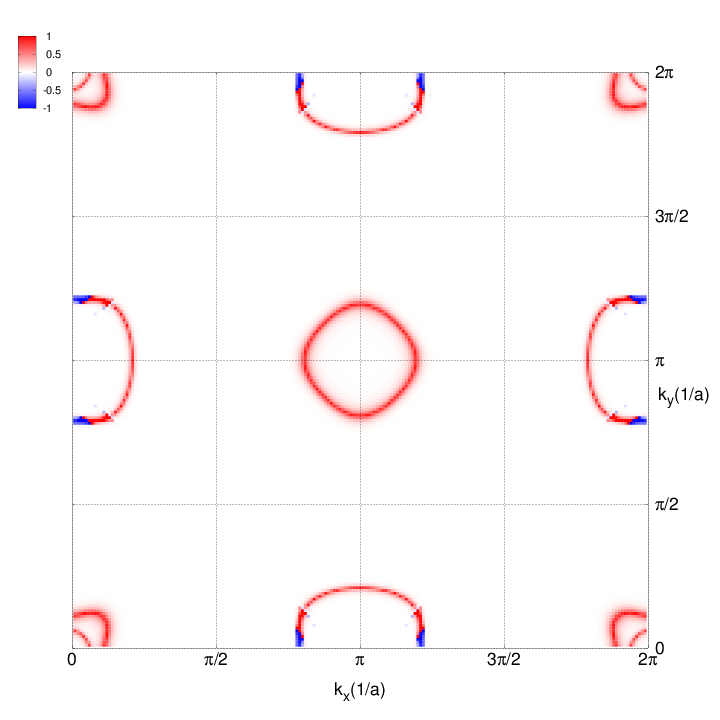}
\caption{Fermi surface projection of $\sum_n \tilde{f}({\bf k})_{n,n}$
  plotted on the unfolded Brillouin zone (one Fe unit cell).}
\label{improper_s_band_space}
\end{figure}

In Fig.~\ref{improper_s_band_space} we plot 
$\tilde{f}({\bf k})_{m,n}$ with ${\bf k}$'s on the Fermi
surface (in the unfolded Brillouin zone), 
which results from the BCS model interpolation on an ultra dense
$200 \times 200$ k-point grid.
The Figure shows that the global symmetry is the A$_{1g}$. The only
sign change in this picture 
occurs within the M electron pockets, already pointed out in previous
works\cite{graser2,kuroki2}. However, once the global symmetry is
fixed, the $\tilde{f}$ components can have in general a non trivial structure,
which is material dependent, due to its multi-band nature. Here,
we find that for the FeSe there are no other sign changes except for
the ones in the M pocket. It is hard to say if this comes from the
improper s-wave symmetry or the sign change is just ``accidental''.
In the band representation, there is not a clear-cut signature of the
improper s-wave components.

In the next section, we are going to analyze what are the symmetry
implications of the improper-s symmetry,
by studying the full pairing function
$F$ in the orbital space and its related BCS gap.

\subsection{k-resolved image of the BCS gap function and physical pairing}

The BCS model derived in the previous Subsection allows not only to
interpolate $f({\bf k})_{\mu,\nu}$ on a dense k-point grid, but also
to evaluate the k-resolved BCS gap $\Delta^{BCS}({\bf k})$, of paramount importance to make a
connection with some angle-resolved experiments as for instance the specific heat
measurements performed in Ref.~\onlinecite{zeng} for the FeSe. That
experiment indicates the presence of gap minima along the $\Gamma$-M
(Fe-Fe) bond directions, and conversely the gap maxima along the
Fe-Se-Fe directions, being the angular dependence fourfold.

\begin{figure}[ht]
\begin{center}$
\begin{array}{cc}
\includegraphics[width=0.5\columnwidth]{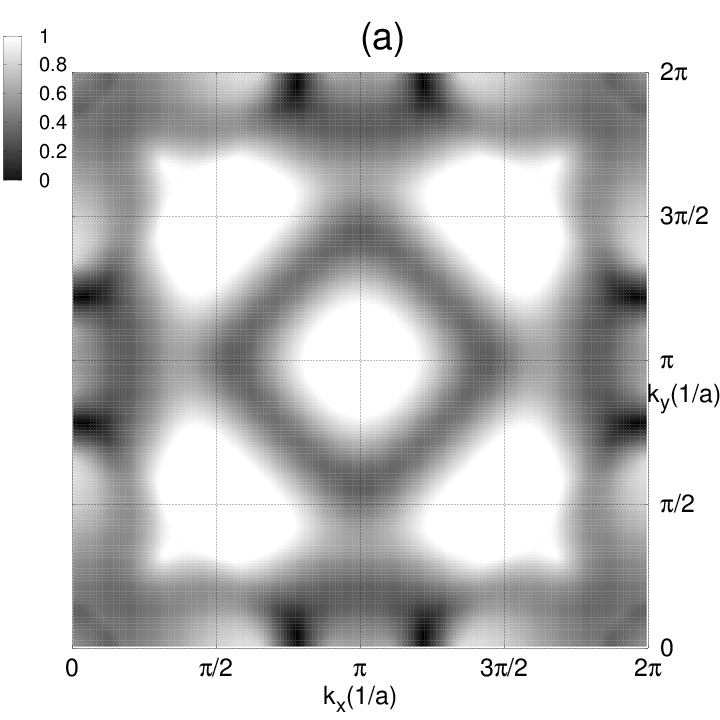} &
\includegraphics[width=0.5\columnwidth]{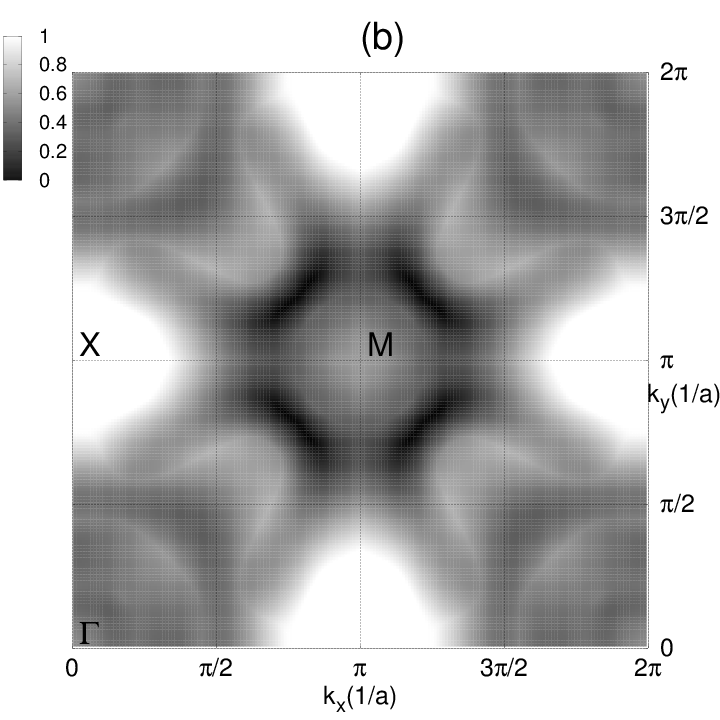}
\end{array}$
\end{center}
\caption{BCS gap computed in eV on a 200 $\times$ 200 k-point grid in the
  unfolded (panel (a)) and folded (panel (b)) Brillouin zone}
\label{bcs_gap}
\end{figure}

Fig.~\ref{bcs_gap} obtained on a $200 \times 200$ k-point grid from
the solution of the BCS model with parameters in
Subsec.~\ref{k-resolved_coupling} shows that indeed minima are aligned
in the $\Gamma$-M directions, while maxima are clearly peaked along
the X-M directions, in general agreement with the specific heat experiment.

\begin{figure}[ht]
\begin{center}$
\begin{array}{cc}
\includegraphics[width=0.5\columnwidth]{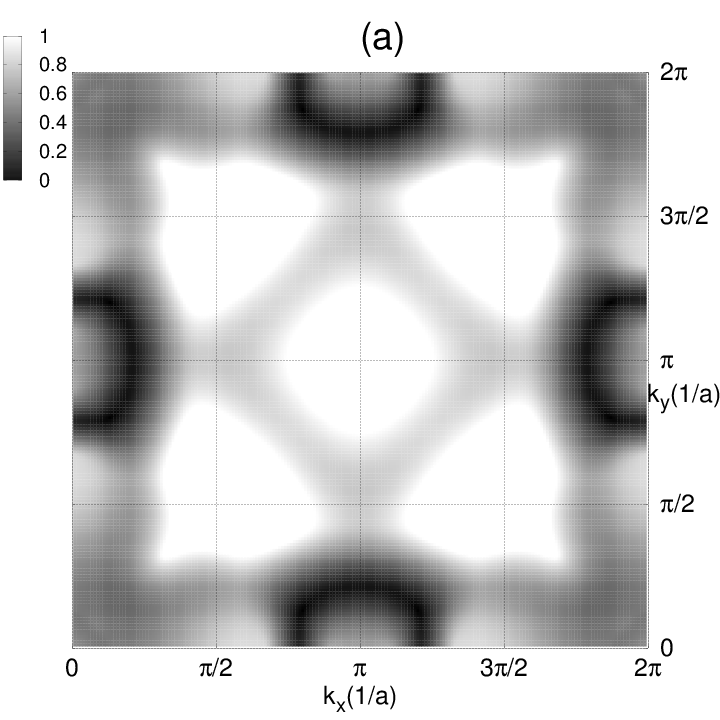} &
\includegraphics[width=0.5\columnwidth]{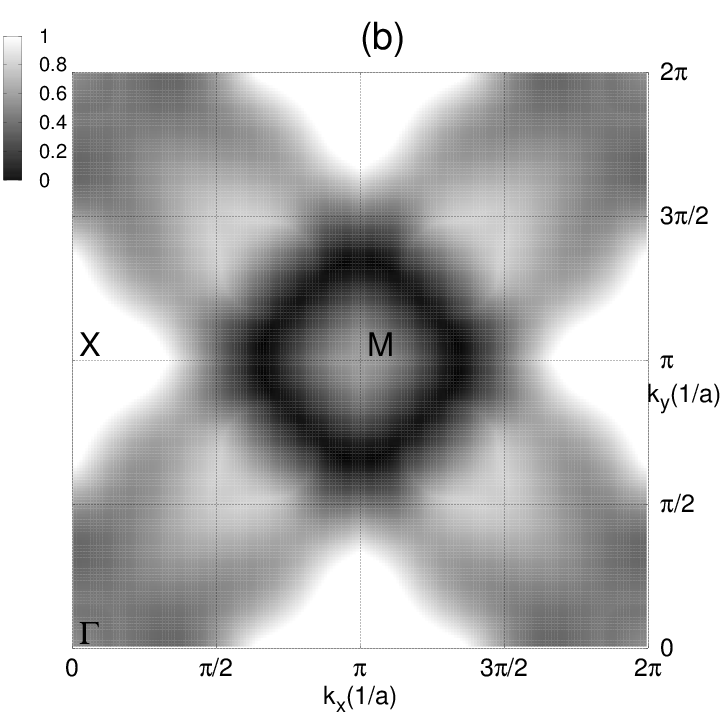}
\end{array}$
\end{center}
\caption{BCS gap computed in eV on a 200 $\times$ 200 k-points grid in the unfolded
 (panel (a)) and folded (panel (b))  Brillouin zone by setting $\Delta_{xz,xy}$ and $\Delta_{yz,xy}$ to zero.}
\label{bcs_gap_noxy}
\end{figure}

To understand the role of the even-odd/odd-even couplings in setting
the gap modulation, we explicitly set to zero $\Delta_{xz,xy}$ and
$\Delta_{yz,xy}$ in Eq.~\ref{delta_bcs_couplings}, while keeping
unchanged the other BCS parameters. The resulting BCS gap of this modified model is
shown in Fig.~\ref{bcs_gap_noxy}. As one can see, the angular
dependence of the BCS gap minima (black regions) of the latter model is
much more isotropic. Thus, the $\Delta_{xz,xy}$ and
$\Delta_{yz,xy}$ BCS parameters, which couple orbitals of different
parity, play a key role to trigger the angular dependence of the BCS gap.

We turn now our attention to the Fourier analysis of the physical
pairing function $F({\bf r},{\bf r}^\prime)=\langle {\bf r}, {\bf r}^\prime| {\hat F}
| 0 \rangle$, where $ {\hat F}$ is the pairing operator in
Eq.~\ref{F}. We express the function $F({\bf r},{\bf r}^\prime)$ in
the orbital basis, and we resolve it in terms of even and odd
components, $F_{ee}$, $F_{eo}$, $F_{oe}$, and $F_{oo}$,  
as in Eq.~\ref{F_components}.

From the their translational properties, one
can easily show that they all depend on the
vector ${\bf r}^\prime - {\bf r}$ connecting the electron pair, and
they are periodic functions (with periodicity set by the lattice
vectors) of the pair center ${\bf R}_\textrm{center}$. The
dependence on ${\bf R}_\textrm{center}$ gives rise to an
inhomogeneous contribution within the unit cell, while the ${\bf
  r}^\prime - {\bf r}$ dependence is set by the symmetry
transformations under the 2D lattice point-group operations.

Therefore, their orbital-summed Fourier decomposition can be written in general as
\begin{equation}
F({\bf r},{\bf r}^\prime) = \sum_{{\bf
  k}, {\bf G}, {\bf G}^\prime} F({\bf k}, {\bf G}, {\bf G}^\prime)
e^{i{\bf k}\cdot({\bf r}^\prime - {\bf r})} e^{-i {\bf G}\cdot {\bf
    r}} e^{-i {\bf G}^\prime\cdot {\bf
    r}^\prime},
\label{Fourier_decomposition}
\end{equation}
where ${\bf k}$ leaves in the 2D Brillouin zone of the iron lattice
plane, and ${\bf G}$ and ${\bf G}^\prime$ are reciprocal lattice
vectors of the 3D Bravais lattice of the crystal.

From Eq.~\ref{Fourier_decomposition}, their parity-resolved Fourier components are
analytically given by
\begin{equation}
F_{\textrm{p}\textrm{p}^\prime}({\bf k}, {\bf G}, {\bf G}^\prime) =  
\sum_{\substack{\{\nu \} \in \textrm{p} \\\{\mu \} \in \textrm{p}^\prime } }
f({\bf k})_{\nu,\mu} 
d_{{\bf k},\nu}({\bf G}) d_{-{\bf k},\mu}({\bf G}^\prime),
\label{Fourier_components}
\end{equation}
where $p$ ($p^\prime$) is the parity under reflection through the iron
plane of the left (right) orbital, and $d_{{\bf k},\nu}({\bf G})$ is the Fourier component of the $\nu$
orbital evaluated at the ${\bf k}+ {\bf G}$ reciprocal
vector in the gauge representation, corresponding to the $d_{{\bf
    k}+{\bf Q},\nu}({\bf G})$ component in the ``physical''
representation. 
Eq.~\ref{Fourier_components} can be evaluated on a dense
k-point grid thanks to the BCS interpolation $f^{BCS}({\bf k})_{\nu,\mu}$ of the QMC pairing
function. In the following we compute Eq.~\ref{Fourier_components} by
taking  $f^{BCS}({\bf k})_{\nu,\mu}$ 
and the Fourier components of the periodic Gaussian d-orbital with
exponent $Z=0.8$.

\begin{figure}[ht]
\begin{center}$
\begin{array}{cc}
\includegraphics[width=0.5\columnwidth]{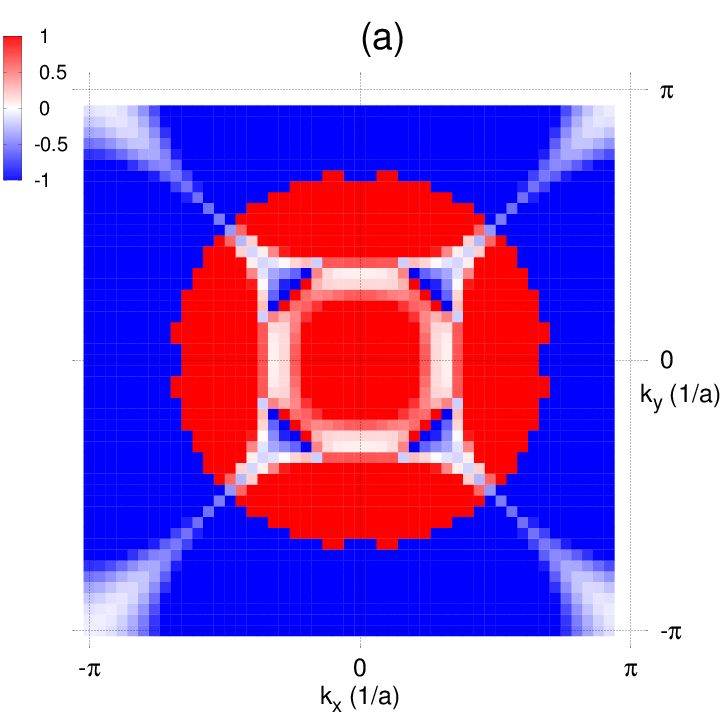} &
\includegraphics[width=0.5\columnwidth]{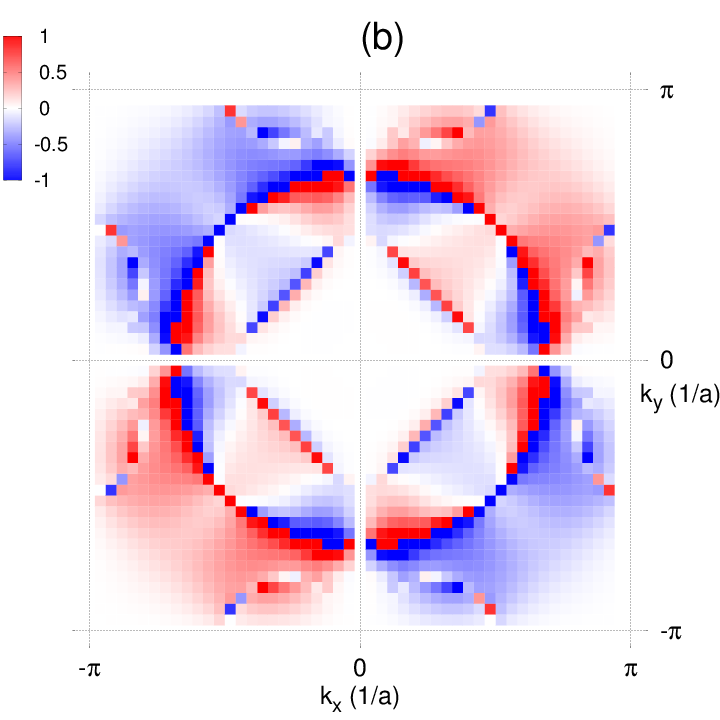}
\\
\includegraphics[width=0.5\columnwidth]{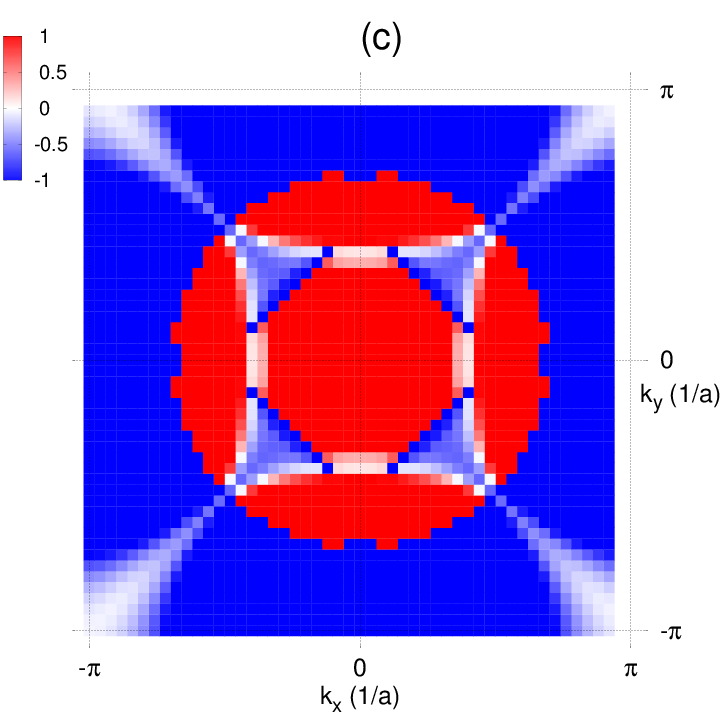} &
\includegraphics[width=0.5\columnwidth]{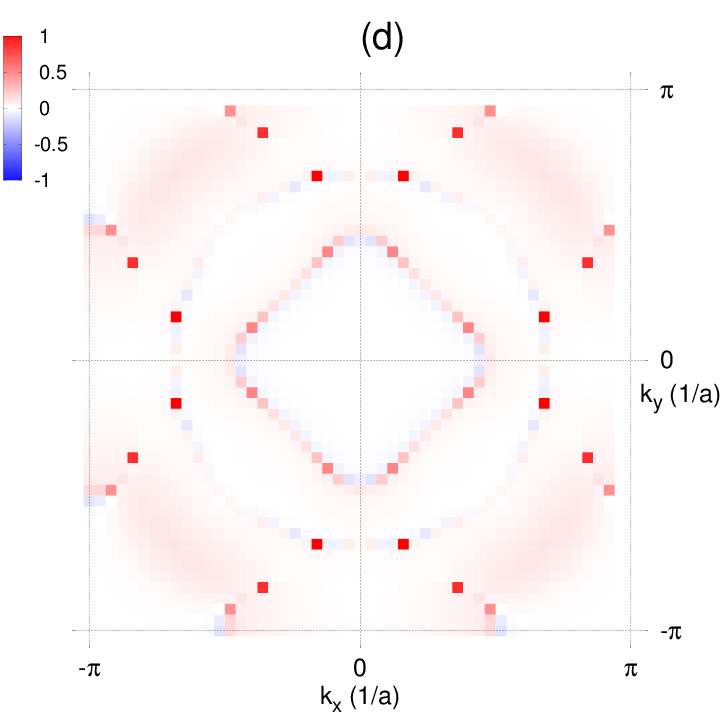}
\end{array}$
\end{center}
\caption{Fourier components of the pairing function on a 50 $\times$
  50 k-point grid in the unfolded Brillouin zone.
Panel (a): even-even $G_z=0$ $G'_z=0$ component;
Panel (b): even-odd (equal to odd-even by symmetry) $G_z=\frac{2
  \pi}{c}$ $G'_z=\frac{2 \pi}{c}$ component;
Panel (c): even-even $G_z=\frac{2 \pi}{c}$ $G'_z=\frac{2 \pi}{c}$;
Panel (d): odd-odd $G_z=\frac{2 \pi}{c}$ $G'_z=\frac{2 \pi}{c}$.}
\label{fourier_components_plot}
\end{figure}

Within the ``improper s-wave'' picture,
$F_{\textrm{p}\textrm{p}^\prime}({\bf k}, {\bf G}, {\bf G}^\prime)$
have the following properties, according to the z-component values of
${\bf G}$ and ${\bf G}^\prime$:
\begin{itemize}
\item For $ {\bf G}_z =0$ and ${\bf G}_z^\prime = 0$, the only non-zero
  components are of type ``even-even'', they are purely real, and they have a standard
  s-wave symmetry, as seen in Fig.~\ref{fourier_components_plot}(a);
\item For $ {\bf G}_z \ne 0$, or  $ {\bf G}_z^\prime \ne 0$, the improper symmetry shows up
  in the ``even-odd'' and ``odd-even'' components, which are purely
  imaginary, and they have a d$_{xy}$-wave symmetry (see, for instance, Fig.~\ref{fourier_components_plot}(b));
\item All the other ``even-even'' and ``odd-odd'' components are purely
  real and of s-wave symmetry (see
  Figs.~\ref{fourier_components_plot}(c) and
  \ref{fourier_components_plot}(d)). 
\end{itemize}

\section{Physical consequences of the improper s-wave symmetry}
\label{physical_consequences}

\begin{figure}[ht]
\includegraphics[width=\columnwidth]{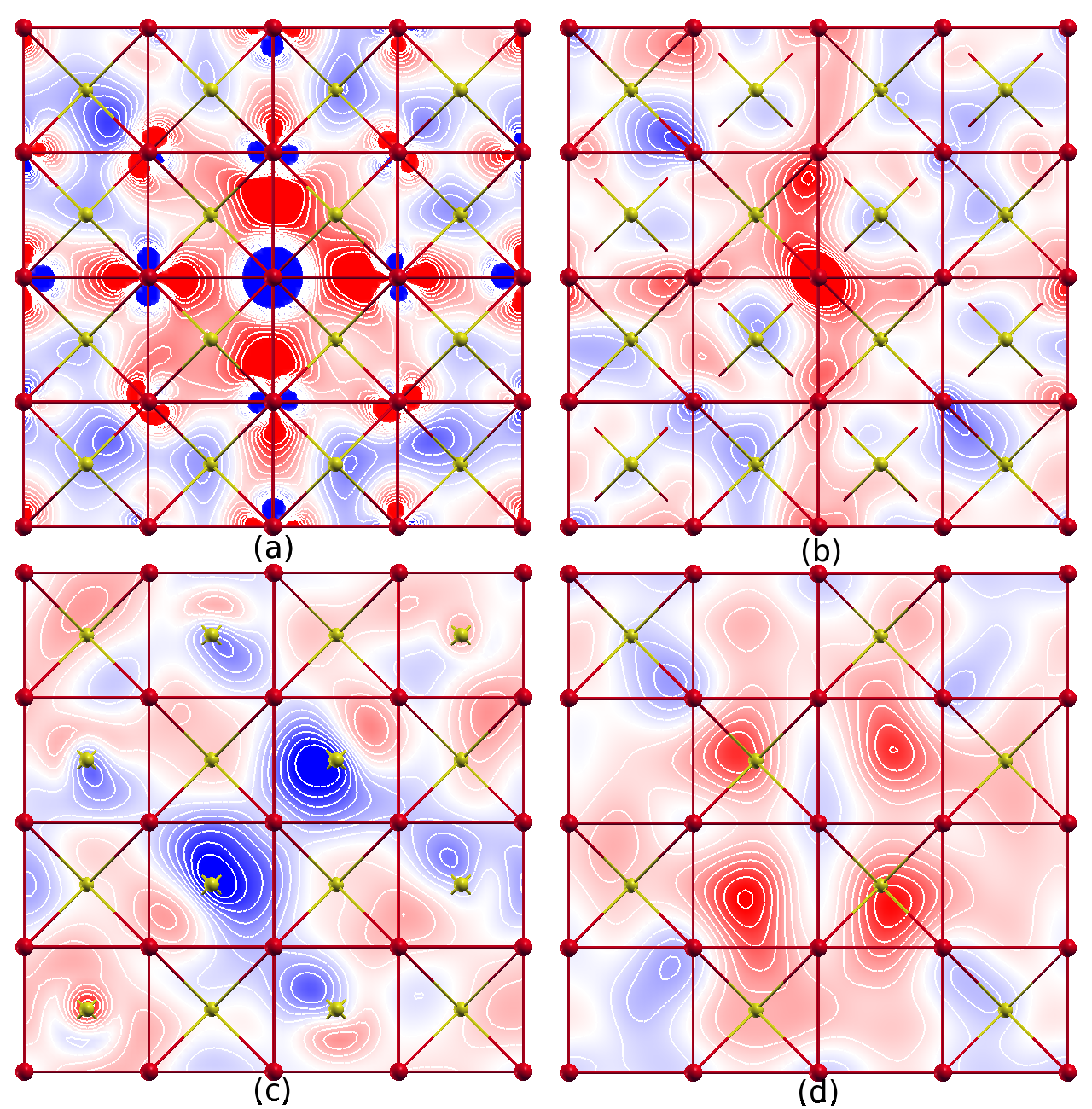}
\caption{Contour plot of the total QMC pairing function projected on
  a plane at different heights in the unit cell. (a) is at $z=0$ which
  coincides with the iron plane, (b) is at $z=0.125 c$, (c) is at
  $z=0.25 c$, and finally (d) is at $z=0.5 c$.}
\label{total_QMC_pairing}
\end{figure}

We explore now the physical consequences of the improper
s-wave from a general
perspective, by assuming that also the other IBS families 
have a $A_{1g}$ superconducting gap, as the FeSe.
First of all, the twofold symmetry of the pairing 
function observed by STM\cite{twofold} 
in stoichiometric FeSe with impurities and magnetic field can be
explained by the Andreev resonance probed by the STM  and sensitive
to the sign change induced by a planar superposition of the
s- and d-wave channels, invariant under $\pi$ rotations. 
Note that the  d$_{xy}$  arrangement is in 
agreement with the different orientations measured by the
STM zero bias conductance.
In Fig.~\ref{total_QMC_pairing}, by plotting the QMC pairing function
at different distances from the iron plane, we show how the $C_4$ is broken as
soon as the experimental local probe is lifted from the iron layer.  

From ab-initio DFT and DMFT
calculations,\cite{aichhorn,werner}
the $d_{x^2-y^2}$ and $d_{3 z^2 -r^2}$ orbitals are usually
pseudogapped.
This allows us to disregard them
from the summation in 
Eq.~\ref{F}.
Thus, $d_{xy}$ is the only
remaining orbital with even parity, 
and 
it can trigger 
the formation of the d-wave channel in the pairing function.
Indeed, for absent or weak $d_{xy}$ spectral weight at the
Fermi level, the two orbitals left in $F$ are both odd under
$\sigma_h$, and therefore they can only contribute to the s-wave channel.

In the ``111'' family, 
DFT band structure calculations and de Haas-van Alphen experiments show
that the main effect of 
P isovalent substitution
is the formation of the third
hole pocket 
- with dominant $d_{xy}$ character -, absent in the
LiFeAs.\cite{putzke} The enhanced 
$d_{xy}$ spectral weight in the P substituted 
compound and the subsequent development of an ``even-odd''
d-wave pairing gap
can explain why the LiFeP pairing is nodal, while the LiFeAs shows a fully gapped
behavior. 

Within the improper s-wave picture, the properties of the ``1111''
family modified by impurity
substitution as in Ref.~\onlinecite{impjap}
are consistent with a 
tendency of $F$  to develop nodes by going from Co to Mn doping
through the opening of the d-wave channel. Indeed, Co impurities are
electron donors, while Mn impurities dope the system with hole, thus
enhancing the $d_{xy}$ spectral weight around $\Gamma$.

The ``122'' family does not
belong to the $P4/nmm$ space-group. 
Nevertheless, by inspection of its $I4/mmm$ space-group unit cell, one
can see that also in that case the $S_4$ improper rotation is the valid symmetry
on the iron layer, thus the improper s-wave theory applies also here.
The $s$-to-$d$ wave transition by doping the BaFe$_2$As$_2$ with K is
one of the most striking
experimental situations for us.\cite{s2d}
In the strong overdoped regime, the
band structure certainly shows all three pockets at $\Gamma$ with
again an enhanced $d_{xy}$  spectral weight. If the d-wave channel
dominates in the improper s-wave gap, then it is plausible that the
experimental thermodynamic 
features are compatible with a pure d-wave behavior. 

Finally, we would like to point
out that the importance of the third hole pocket, with strong $d_{xy}$
character, 
 has already been
observed by Kuroki and co-workers in relation to the
 $T_c$ dependence on the pnictogen height.\cite{kuroki}
The improper s-wave symmetry
explains 
the key role played by $d_{xy}$, particularly in connection
with the appearance of gap nodes, the twofold anisotropy, and the
stabilization of the 
d-wave channel, by reconciling a series of experiments which look
otherwise contradictory. It suggests also that the
pnictogen or the chalcogen, rather than being spectators, take an
active role in the electronic pairing, by bridging two next-nearest
neighbor iron sites in a $d_{xy}$ fashion via the improper s-wave symmetry.

From the theoretical point of view, there is a lot of confusion in literature.
For instance, to our knowledge, no RPA and no mean-field studies have been done so
far with a low energy 
2D model explicitly compatible with the improper rotations and
gauged translation invariance.
As we have shown translation invariance 
and $C_{4\nu}$ have to be combined in an appropriate way, by taking particular attention to the 
even/odd $\sigma_h$  reflection of the orbitals in the model.
In our QMC simulations the improper s-wave symmetry is stabilized,
which determines a highly entangled nodal structure of the pairing function, 
and the 
possibility for mixing between s-wave and d-wave components, an effect that, as we 
discussed, is crucial to understand most experiments in the iron-based superconductors. 

\begin{acknowledgments}
We acknowledge useful discussions with S. Biermann, T. Cren,
M. Fabrizio, A. Savin, and D. J. Scalapino. 
The HPC resources have been provided under the IDRIS/GENCI 2012096493 and CINECA-MIUR
ISCRA-HP10A2GZHV grants.
\end{acknowledgments}


\begin{thebibliography}{99}
\bibitem{conventions} In the following, for simplicity of notations, 
 we identify the subgroup symbols $S_4$ or $C_4$ with the 
elementary operation (e.g. rotation by $\pi/2$) that defines the cyclic subgroup.
\bibitem{footnote_iron_center} Strictly speaking, the $d$ orbitals are
  zero at the iron sites. The amplitude of the pairing function is
  computed by integrating those orbitals over a planar mesh
  ($+\delta_x,-\delta_x,+\delta_y,-\delta_y$, with $|\delta|=0.25$ Bohr) around the
  iron center. The sign of the integrand on the mesh selects their angular symmetry.
\bibitem{footnote_pwscf} The LDA-DFT calculations have been performed
  with the {\textsc quantum-espresso}\cite{qe}  code. Fe and Se atoms are 
described respectively by ultrasoft (with 16 valence electrons) and
norm-conserving (with 6 valence electrons) pseudopotentials generated
with a scalar-relativistic calculation. The plane-wave (PW) cutoff is
60 Ry for the wave-function, and 240 Ry for the charge.  A $4 \times 4 \times 4$
electron-momentum grid and a Methfessel-Paxton smearing  of 0.01 Ry are
used in the electronic integration. 

Wannierization has been performed with the {\textsc Wannier90}\cite{wannier90}  
program on a $N_w = 4\times4\times4$
electron-momentum mesh, by including the d states on the two
inequivalent Fe site of the unit cell.
The localized orbitals yield Wannierized bands in a very good agreement with
those computed in the PW basis set
in a window of $\pm$ 1.5 eV around the Fermi level.
\bibitem{foot_temperature} 
At ambient pressure,
 the slightly broken low temperature phase is
  very close 
  to the tetragonal parent structure, and their difference  (around
  0.01\AA~between the $a$ and
  $b$ axes) has been
  neglected in our calculations. At 4 GPa, the $a/b$
  anisotropy is instead more than four times stronger, and thus we used
  the corresponding experimental low temperature orthorombic structure.
\end{thebibliography}
\end{document}